\documentclass[]{interact}

\usepackage{gensymb}
\usepackage{epstopdf}
\usepackage[caption=false]{subfig}

\usepackage{natbib}
\bibpunct[, ]{(}{)}{;}{a}{}{,}

\theoremstyle{plain}

\theoremstyle{definition}

\theoremstyle{remark}

\begin{document}

\articletype{RESEARCH ARTICLE}

\title{Shoulder Physiotherapy Exercise Recognition: Machine Learning the Inertial Signals from a Smartwatch}

\author{
\name{David M. Burns MD\textsuperscript{a}\thanks{CONTACT David M. Burns. Email: d.burns@utoronto.ca}, Nathan Leung\textsuperscript{b}, Michael Hardisty PhD\textsuperscript{c}, Cari Whyne PhD\textsuperscript{c}, Patrick Henry MD FRCSC PhD\textsuperscript{a,d}, and Stewart McLachlin PhD\textsuperscript{b}}\affil{\textsuperscript{a}Division of Orthopaedic Surgery, University of Toronto, Canada \textsuperscript{b}Dept of Mechanical \& Mechatronics Engineering, University of Waterloo, Canada \textsuperscript{c} Sunnybrook Research Institute, Toronto, Canada \textsuperscript{d} Sunnybrook Health Sciences Centre, Toronto, Canada}}
\maketitle

\begin{abstract}
\textbf{Objective: } 
Participation in a physical therapy program is considered one of the greatest predictors of successful conservative management of common shoulder disorders. However, adherence to these protocols is often poor and typically worse for unsupervised home exercise programs. Currently,  there are limited tools available for objective measurement of adherence in the home setting. The goal of this study was to develop and evaluate the potential for performing home shoulder physiotherapy monitoring using a commercial smartwatch. \\
\textbf{Approach: } 
Twenty healthy adult subjects with no prior shoulder disorders performed seven exercises from an evidence-based rotator cuff physiotherapy protocol, while 6-axis inertial sensor data was collected from the active extremity. Within an activity recognition chain (ARC) framework, four supervised learning algorithms were trained and optimized to classify the exercises: k-nearest neighbor (k-NN), random forest (RF), support vector machine classifier (SVC), and a convolutional recurrent neural network (CRNN). Algorithm performance was evaluated using 5-fold cross-validation stratified first temporally and then by subject.  \\ 
\textbf{Main Results: } 
Categorical classification accuracy was above 94\% for all algorithms on the temporally stratified cross validation, with the best performance achieved by the CRNN algorithm (99.4\%). The subject stratified cross validation, which evaluated classifier performance on unseen subjects, yielded lower accuracies scores again with CRNN performing best (88.9\%). \\ 
\textbf{Significance:  }
This proof of concept study demonstrates the technical feasibility of a smartwatch device and supervised machine learning approach to more easily monitor and assess the at-home adherence of shoulder physiotherapy exercise protocols. 

\end{abstract}


\section{Introduction}

Symptomatic degenerative rotator cuff (RTC) tears are one of the most common causes of shoulder pain and dysfunction \citep{mitchell_shoulder_2005}. Conservative management with physical therapy has been established as an effective first line treatment for this condition, yielding significant improvements in patient reported outcome scores and obviating the need for surgery in the majority of patients \citep{kuhn_effectiveness_2013}. Participation in a physical therapy program is considered one of the greatest predictors of successful conservative management, however, adherence to standard exercise protocols is often poor (around 50\%) and typically worse for unsupervised home exercise programs \citep{holden_recommendations_2014}. Objective measures of adherence are therefore important from a clinical standpoint, but measures to monitor adherence would also be useful to evaluate both the participation as well as the success of any given physical therapy protocol. 

Currently, measures of adherence are often lacking from clinical trials of physical therapy for musculoskeletal conditions \citep{holden_recommendations_2014}. When adherence is monitored, investigators typically utilize non-objective measures such as physiotherapy clinic attendance and patient self-reports for home-exercises, which are of questionable validity \citep{jack_barriers_2010}. An alternative approach that has yet to be explored in the setting of shoulder therapy is to directly measure patient adherence. 

Advances in the quality and capabilities of embedded inertial sensors within smartwatches and machine learning algorithms present an opportunity to leverage these technologies and enable objective measurement of home shoulder physiotherapy adherence. Toward this end, the purpose of this study is to assess the feasibility of performing shoulder physiotherapy exercise recognition with inertial sensor data recorded from a wrist-worn device. 

\subsection{Hypothesis}

It was hypothesized that shoulder physiotherapy exercises could be classified from the temporal sequence of inertial sensor outputs from a smartwatch worn on the extremity performing the exercise.

\subsection{Related Work}

A great deal of active research is being conducted in the HAR field with inertial sensor data, and the reviews articles by \cite{attal_physical_2015} and \cite{bulling_tutorial_2014} provide an excellent summary. HAR studies using wrist worn or smartwatch based inertial sensors have been performed \citep{nguyen_activity_2015,garcia-ceja_long-term_2014,yang_using_2008}, however, these authors address classification of activities of daily living. Inertial sensors have also been studied for use in clinical shoulder evaluation \citep{de_baets_shoulder_2017,korver_inertia_2014,pichonnaz_enhancing_2015}, kinematic analysis \citep{crabolu_vivo_2017,picerno_ambulatory_2015}, range of motion measurement \citep{werner_validation_2014}, and upper extremity pose estimation \citep{shen_i_2016}. To our knowledge, this study is the first to apply machine learning to wrist worn inertial sensor data for shoulder physiotherapy exercise recognition. 

\section{Materials and Methods}

\subsection{Dataset}

Twenty healthy adult subjects with asymptomatic shoulders and no prior shoulder surgery were recruited and provided informed consent for participation in this study. The subjects' mean age was 28.9, range 19-56. There were 14 male and 6 female subjects. Fifteen subjects were right hand dominant, and five were left-hand dominant. Two subjects had a prior low-grade acromio-clavicular joint separation, and one subject had a prior clavicle fracture.

Under the supervision of an orthopaedic surgeon, each subject performed 20 repetitions of seven shoulder exercises bilaterally. The exercises performed are elements of an evidence-based rehabilitation protocol for full-thickness atraumatic rotator cuff tears \citep{kuhn_effectiveness_2013} and included: pendulum (PEN), abduction (ABD), forward elevation (FEL), internal rotation (IR), external rotation (ER), trapezius extension (TRAP), and upright row (ROW). IR, ER, and TRAP exercises were performed with a medium resistance band, and ROW was performed with a 3 lb weight. Videos demonstrating the exercises are included in the data supplement. 

6-axis (acceleration and gyroscope) inertial sensor data was acquired from the active extremity using an Apple Watch (Series 2 \& 3) with the PowerSense app, sampling at $f_s = 50$ Hz.

\subsubsection{Ground Truth Annotation}

Labeling of the raw data was performed during acquisition by collecting separate data files corresponding to each exercise set. The labeling was refined using a semi-autonomous algorithm to label the active portion from each file, by thresholding the $T=2$ second moving average $\overline{e}_a$ of the accelerometer signal energy.
\begin{equation}
\overline{e}_a(t) = \frac{1}{T} \sum_{t=0}^T |\mathbf{a}(t)|^2
\end{equation}

The labeling was visualized and corrected manually as required when the semi-autonomous algorithm failed to produce an appropriate labeling. A sample plot of the raw data is included in the Appendix Figure \ref{f:seg}.

\subsection{Activity Recognition Chain}

\begin{figure}[]
  \begin{center}
    \includegraphics[scale=0.5]{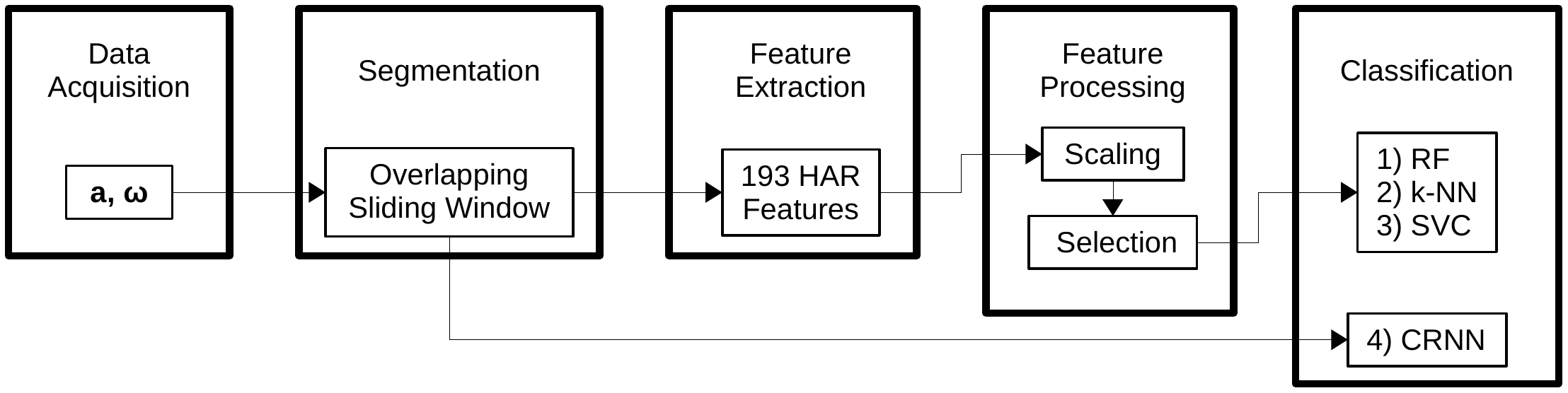}
  \end{center}
  \caption{Activity Recognition Chain. Arrows indicate computational flow.}
  \label{f:arc}
\end{figure}

The approach to activity classification in this work follows the activity recognition chain (ARC) framework proposed by \cite{bulling_tutorial_2014}, and is depicted in Figure \ref{f:arc}. Detailed description of the design and evaluation of each stage in the ARC follow.

\subsubsection{Data Acquisition}

The 6-axis raw sensor data $D$ consists of total acceleration $\mathbf{a} = [a_x, a_y, a_z]$ and rotational velocity $\mathbf{\omega} = [\omega_x, \omega_y, \omega_z]$, measured in the coordinate frame of the watch. No further preprocessing or filtering was applied to the raw data.

\subsubsection{Data Segmentation}

The raw sensor data was segmented using overlapping fixed-length sliding windows $W$ for each of the six sensor signals. The 3D temporal signal tensor $\varphi$ is produced from the set of windows 
\begin{gather}
\varphi = \{W_1,...,W_N\} \\
W_i = \{\mathbf{a_x},\mathbf{a_y},\mathbf{a_z},\mathbf{\omega_x}, \mathbf{\omega_y},\mathbf{\omega_z} \}_i
\end{gather}
and has shape $(N,\frac{L}{f_s}, 6)$, where $L$ is the window length. An exercise label was attributed to a window from the ground truth annotation when the exercise was performed for the entirety of that window. As depicted in Figure \ref{f:arc}, the convolutional recurrent neural network (CRNN) classifier was trained to learn these segments directly. 

\subsubsection{Feature Extraction}

A feature mapping $\mathcal{F}(W)$ comprised of typical HAR statistical and heuristic features \citep{gonzalez_features_2015} was computed to define the feature space for the RF, k-NN, and SVC classifiers. The feature set was comprised of univariate and bivariate features. 

An identical set univariate features: mean, variance ($\sigma^2$), standard deviation ($\sigma$), maximum, minimum, skewness, kurtosis, mean crossings ($\zeta$), mean spectral energy ($\xi$), and a 4-bin histogram, were computed for each signal vector in the segment, and also for the two energy vectors  $\mathbf{e_a}$ and $\mathbf{e_\omega}$
\begin{gather}
\mathbf{e_a} = \mathbf{a_x}^2 + \mathbf{a_y}^2 + \mathbf{a_z}^2 \\
\mathbf{e_\omega} = \mathbf{\omega_x}^2 + \mathbf{\omega_y}^2 + \mathbf{\omega_z}^2
\end{gather}
The mean spectral energy of a vector $\mathbf{x}$ is defined from the fourier transform $f$ as 
\begin{equation}
\xi(\mathbf{x}) = \mathtt{mean}|f(\mathbf{x})|^2
\end{equation}

Pearson correlation coefficients (the bivariate features) were computed between all vector pair combinations. The active extremity side (right or left) was input as the final feature, yielding 133 total features for each instance in the feature space corresponding to a window segment. 

\subsubsection{Feature Processing}

Features were scaled to zero mean and norm standard deviation, and a Gini importance ranking \citep{louppe_understanding_2013} feature selection strategy was implemented to reduce the feature space. The Gini importance feature rankings were computed using an Extremely Randomized Trees classifier \citep{geurts_extremely_2006} with an ensemble of 250 trees, and used to select the highest ranking features for inclusion in the classifier pipeline. 

\subsubsection{Classification}

Four classifiers were selected for evaluation in the ARC, based on prior successful performance for human activity recognition with inertial sensor data sets \citep{bulling_tutorial_2014, attal_physical_2015}. A random forest (RF), k-nearest neighbor (k-NN), and support vector machine classifier (SVC) were trained and evaluated on the feature data, utilizing the implementations provided by the sklearn library \citep{pedregosa_scikit-learn:_2011}. 

\begin{figure}[!htb]
\centering
\subfloat[CRNN architecture and layer output dimensions for 2s sliding window.]{%
\resizebox*{7cm}{!}{\includegraphics{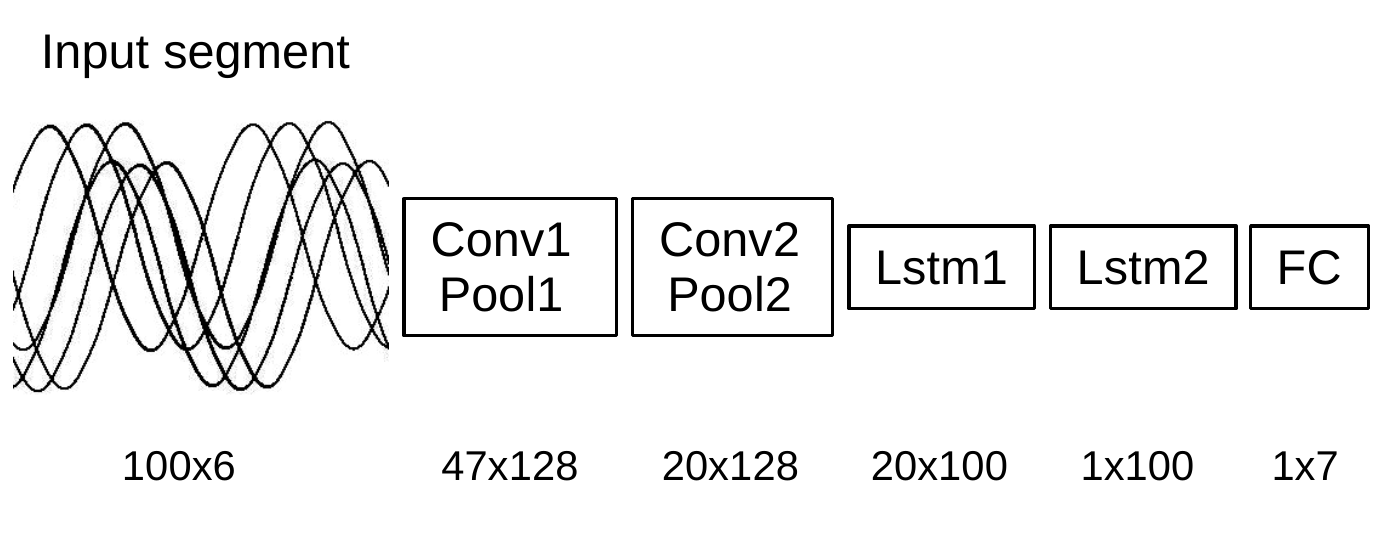}}}\hspace{5pt}
\subfloat[CRNN training curves.]{%
\resizebox*{7cm}{!}{\includegraphics{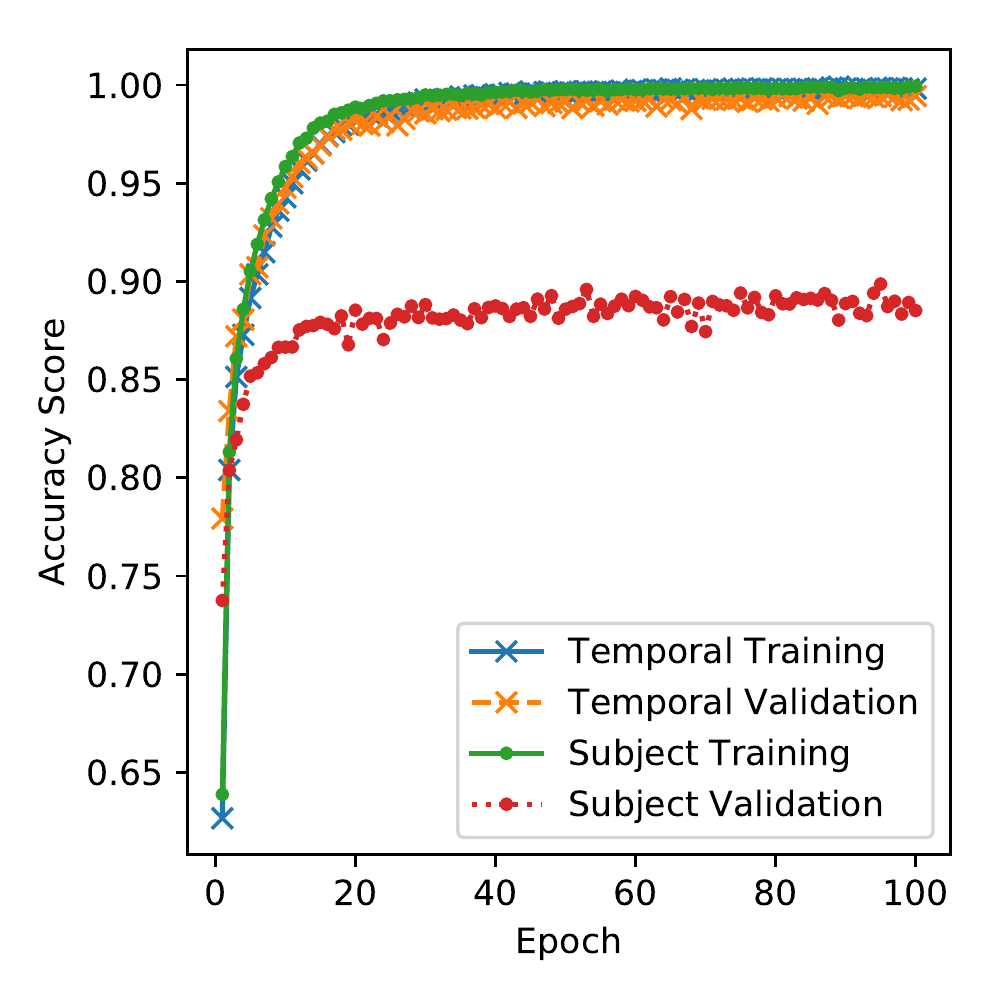}}}
\caption{Convolution layers utilized a kernel size of 7, stride 1, 128 filters, relu activation, and were followed by size 2 max pooling. LSTM layers were regularized with a dropout of 0.1, and utilized tanh and hard sigmoid activation and recurrent activation respectively. The fully connected (FC) layer utilized softmax activation and l2 regularization weighted with $\lambda=0.01$.} \label{f:nnet2}
\end{figure}

A convolutional recurrent neural network (CRNN) was also trained to learn the segments directly using the keras library \citep{chollet_keras_2015}. CRNN conceptually utilizes convolutional input layers to generate a feature representation of the signal tensor which is learned by the recurrent layers utilizing long short term memory (LSTM) cells, an approach utilized recently for video \citep{yue-hei_ng_beyond_2015} and inertial signal \citep{ordonez_deep_2016} classification. Our CRNN, depicted in Figure \ref{f:nnet2}, utilizes two convolutional layers feeding two stacked LSTM layers and a dense classification layer. The network was trained over 100 epochs with the adam optimizer, using a batch size of 1024 on a NVIDIA Tesla K40C GPU. 

\subsubsection{Evaluation}

Parameter tuning and classifier performance was evaluated using 5-fold cross validation with two distinct data splitting approaches. 

Temporal stratified splitting was implemented to assess ARC performance including all subjects in the training set. The temporal stratified folds were calculated with no temporal overlap of segments across folds. 

Subject stratified splitting evaluates ARC performance on subjects outside the training set, and was implemented such that each fold contained only data from four participants unique to that fold.

\section{Results}

\begin{figure}
\centering
\subfloat[CRNN subject stratified confusion matrix.]{%
\resizebox*{6cm}{!}{\includegraphics{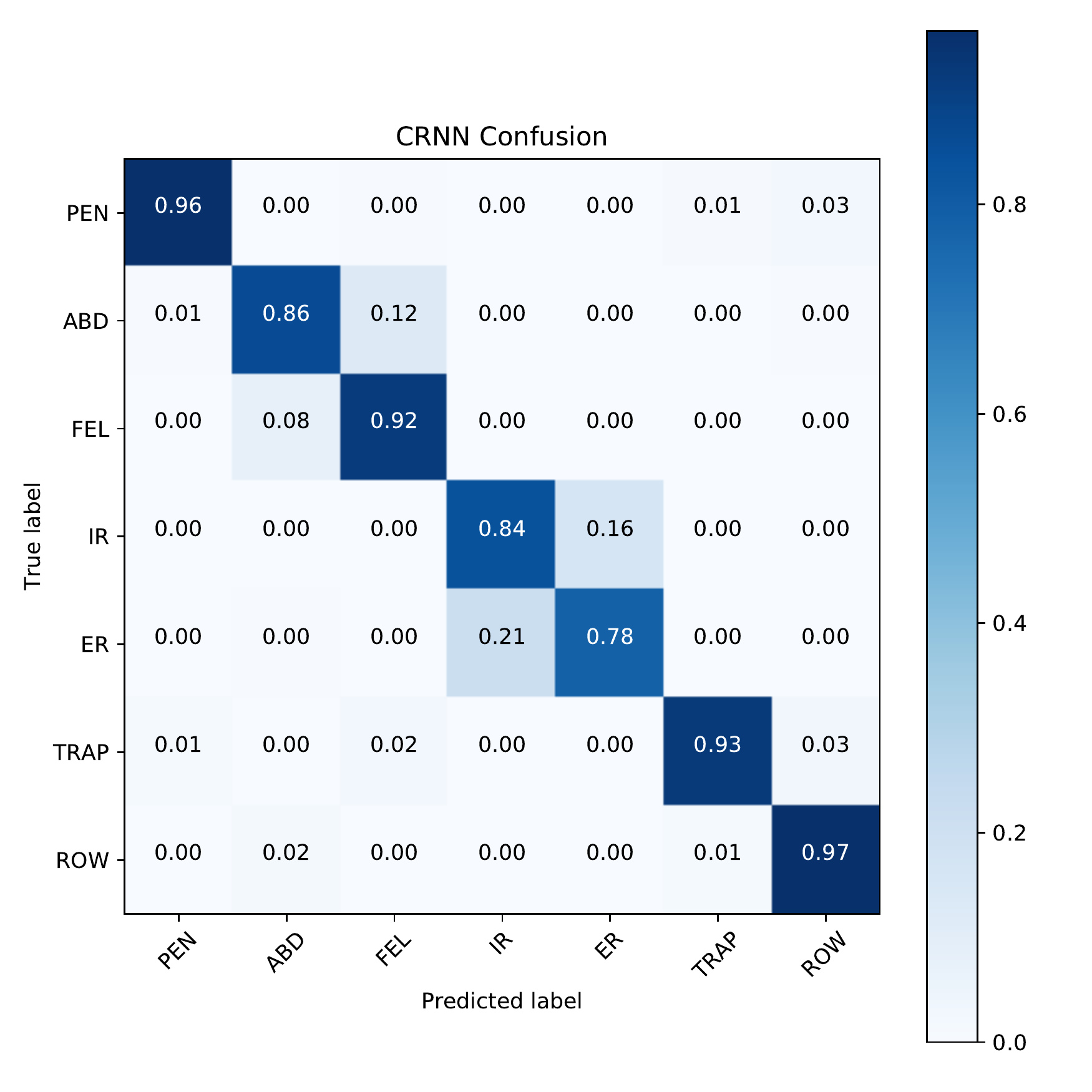}}}\hspace{6pt}
\subfloat[LDA 2-component feature space embedding.]{%
\resizebox*{6cm}{!}{\includegraphics{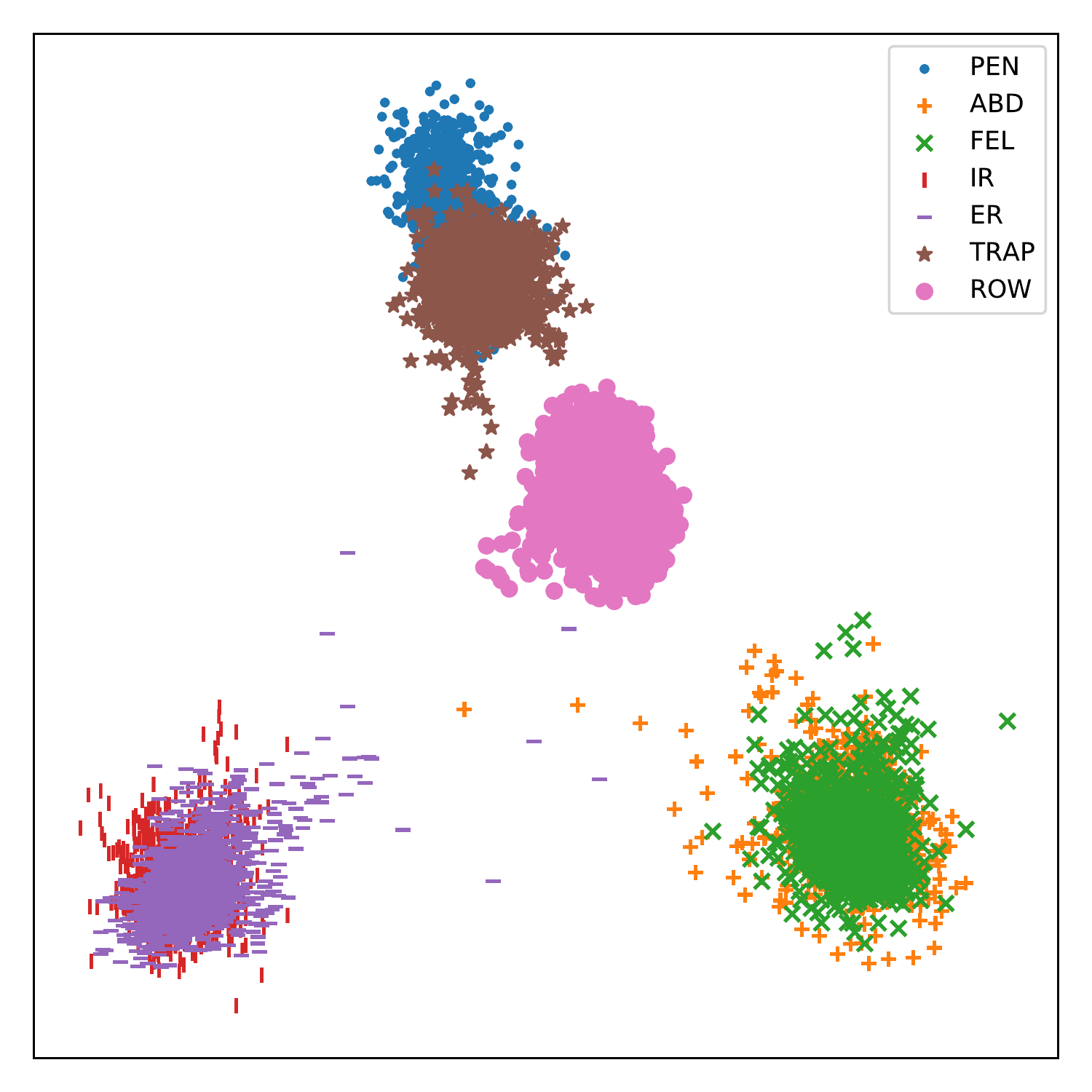}}}
\caption{Similarity of ABD vs FEL and IR vs ER demonstrated by a) confusion matrix and b) linear discriminant analysis (LDA)} \label{f:conf}
\end{figure}

Tuned classifier performance for a two second window, 75\% overlap segmentation is listed in Table \ref{t:perf}. This segmentation strategy yielded 22,053 and 23,397 samples with approximately equal class balance for temporal and subject stratified 5-fold cross validations respectively. For the RF, SVC, and k-NN feature classifiers, the top 75\% Gini ranked features were included in the classification pipeline. A cross-validated confusion matrix is plotted in Figure \ref{f:conf} for the CRNN classifier on the subject stratified data, demonstrating the confusion between ABD vs FEL, and IR vs ER respectively. The similarity of the confused classes in the feature space is also visualized using a 2-component linear discriminant analysis embedding. 

\begin{table}[]
\tbl{Activity recognition chain performance using 2s sliding window and 75\% training overlap.}
{\begin{tabular}{lccccc} \toprule
                     & \textbf{Training} & \textbf{Temporal Validation} & \textbf{Subject Validation} & \textbf{Train Time {[}s{]}} & \textbf{Score Time {[}s{]}} \\ \midrule
\textbf{RF}          & 1.000 $\pm$ 0.000   & 0.975 $\pm$ 0.001              & 0.841 $\pm$ 0.028             & 23.4                        & 0.1                        \\
\textbf{SVC}         & 0.955 $\pm$ 0.003   & 0.941 $\pm$ 0.001              & 0.853 $\pm$ 0.026             & 20.4                        & 3.5                        \\
\textbf{k-NN (k=1)}  & 1.000 $\pm$ 0.000   & 0.978 $\pm$ 0.003              & 0.806 $\pm$ 0.012             & 8.6                         & 15.6                         \\
\textbf{k-NN (k=30)} & 0.964 $\pm$ 0.001   & 0.957 $\pm$ 0.002              & 0.827 $\pm$ 0.016             & 8.2                         & 23.9                         \\
\textbf{CRNN}        & 0.999 $\pm$ 0.001   & 0.994 $\pm$ 0.002              & 0.889 $\pm$ 0.016             & 332.8                       & 7.0                       \\ \bottomrule
\end{tabular}}
\label{t:perf}
\end{table}

\subsection{Segmentation}

The effect of window length and training overlap on classification accuracy are plotted in Figure \ref{f:seg}. As expected increased window length improved performance, however, high accuracy was also achievable with shorter windows ($\leq$1s). Performance was also substantially improved by augmenting the training data with high window overlap. The subjects in this study performed single exercise repetitions at a mean rate of one every 2.2 seconds, range 0.9 to 4.3 seconds. The highest accuracy achieved using 4 second windows and 95\% training overlap was with the CRNN classifier at $99.996 \pm 0.008\,\%$ and $91.2 \pm 0.2\,\%$ for temporal and subject stratified cross validation respectively. 

\begin{figure}
\centering
\subfloat[Segmentation window effect.]{%
\resizebox*{7cm}{!}{\includegraphics{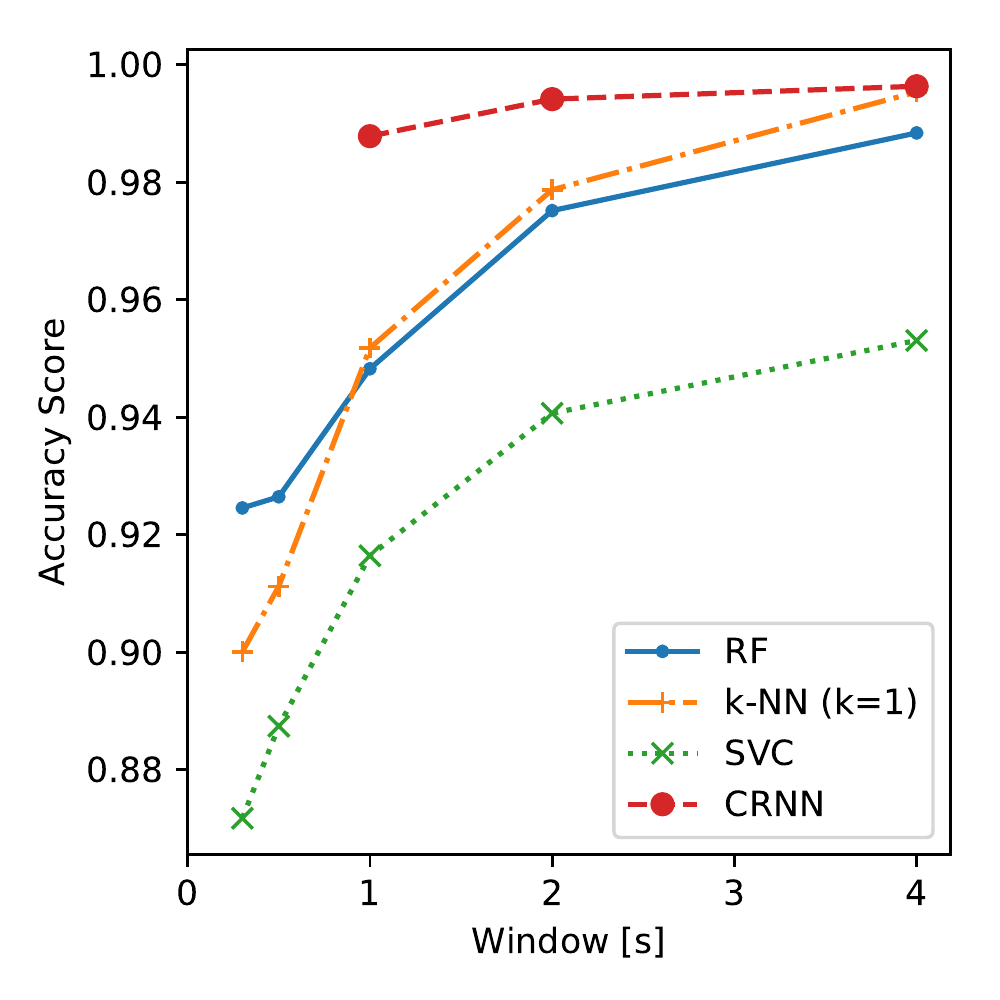}}}\hspace{5pt}
\subfloat[Segmentation overlap effect.]{%
\resizebox*{7cm}{!}{\includegraphics{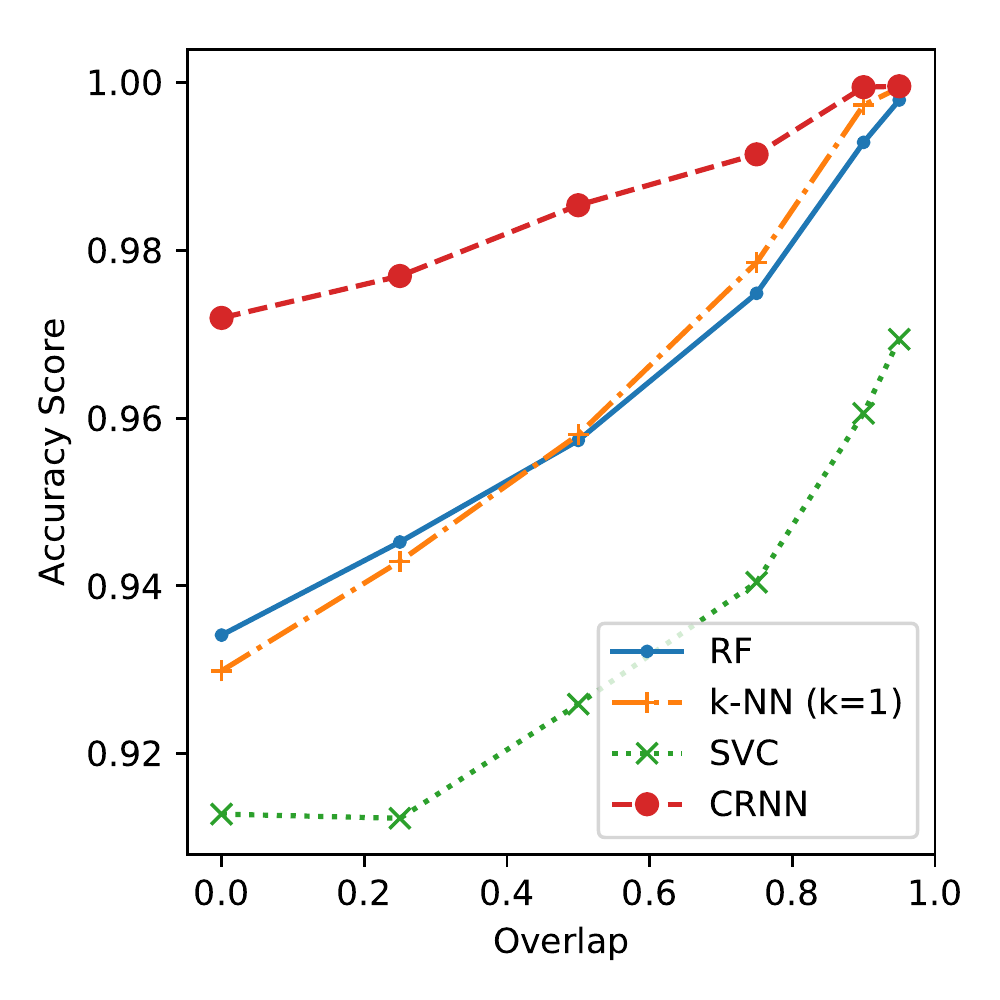}}}
\caption{Segmentation effects on ARC performance with temporal stratified data} \label{f:seg}
\end{figure}

\subsection{Feature Extraction and Selection}

\begin{figure}
\centering
\subfloat[Sorted feature mean Gini importance with standard deviation (uncertainty).]{%
\resizebox*{7cm}{!}{\includegraphics{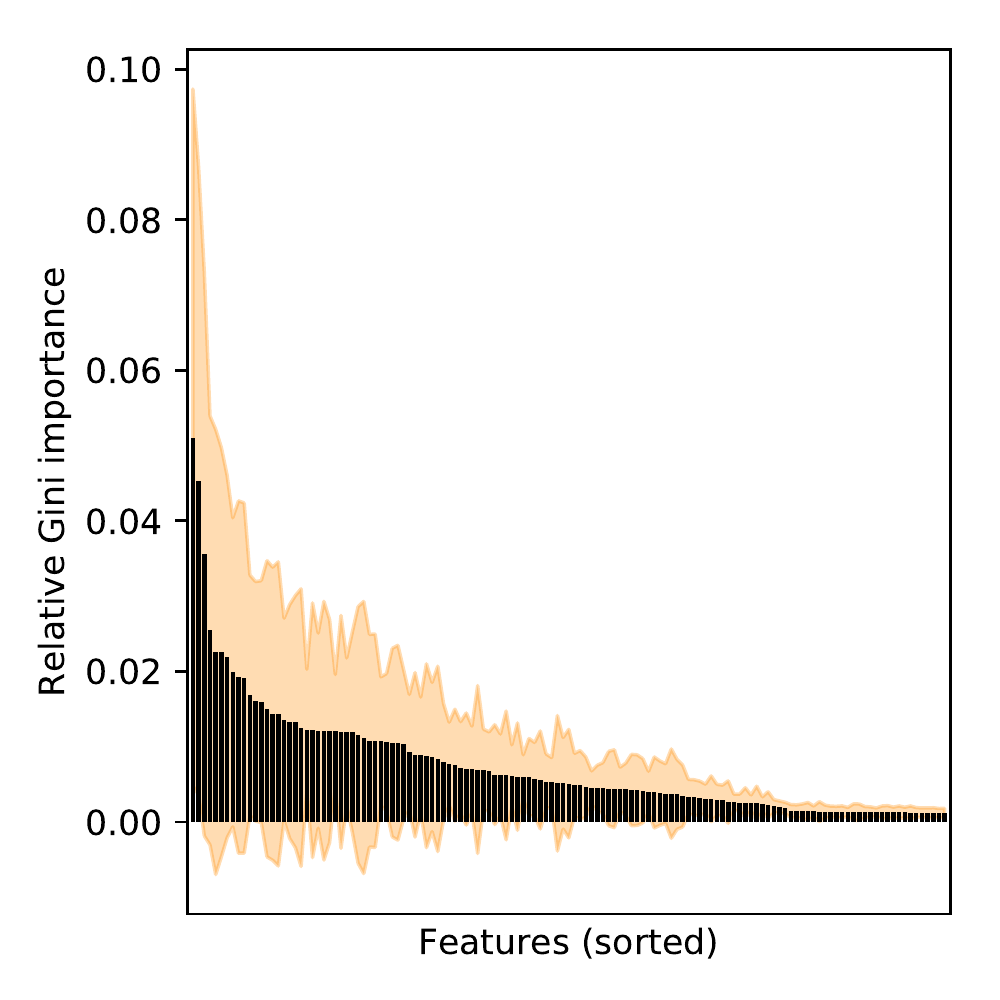}}}\hspace{5pt}
\subfloat[Effect of feature selection on ARC performance.]{%
\resizebox*{7cm}{!}{\includegraphics{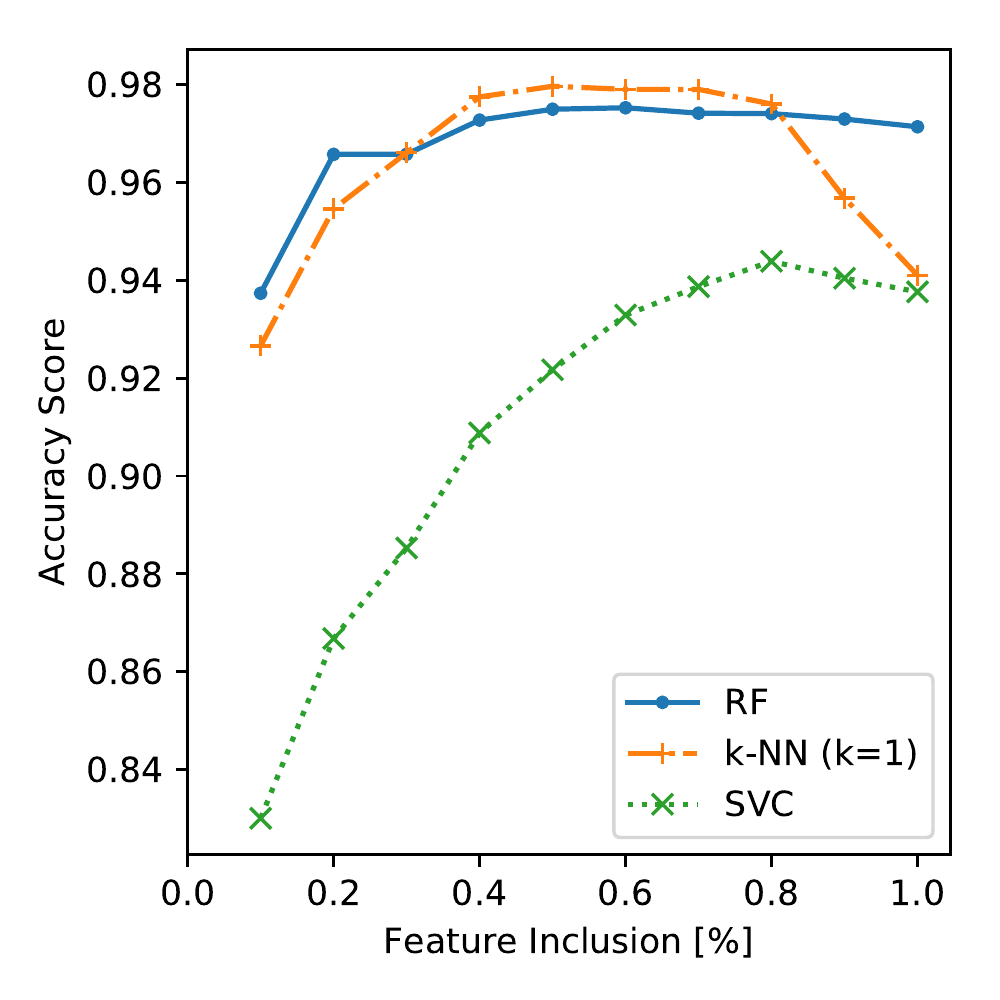}}}
\caption{ARC feature selection} \label{f:select}
\end{figure}

The distribution of feature Gini importance is plotted in Figure \ref{f:select}. The five top ranked features included both time and frequency domain accelerometer features: mean$(a_y)$, $\xi(a_y)$, $\sigma(a_x)$, $\xi(a_x)$, and $\zeta(e_a)$. The Pearson correlation features performed the worst, occupying the bottom 27 of 133 ranks. Classifier performance was optimal using approximately 75\% of the computed features as shown in Figure \ref{f:select}. For the RF and k-NN classifiers, performance was well preserved even down to 10\% feature inclusion.

\subsection{Classifier Optimizations}

Selected hyperparameter optimizations are plotted in Figure \ref{f:hopt} in the Appendix. For the k-NN classifier, $k=1$ and $k=30$ were optimal on the temporal and subject stratified cross validations respectively. For the RF classifier, 150 estimators yielded efficient and accurate results, with optimal performance achieved splitting nodes with maximum 10\% of the feature set. For the SVC classifier, the kernel coefficient $\gamma$ was optimal at 0.005. 

A restricted optimization was performed of the CRNN architecture and hyperparameters including network depth, convolutional kernel size and units, pool size, lstm units, and regularization strategy. Optimal performance was achieved on our dataset using the architecture depicted in Figure \ref{f:nnet2}. 

\section{Discussion}
This study demonstrates that robust and highly accurate classification of shoulder physiotherapy activities from 6-axis inertial sensor data is feasible. The classification accuracy achieved on the temporal stratified data exceeds state-of-the-art performance achieved in other studies using wrist worn sensor HAR datasets \citep{nguyen_activity_2015,garcia-ceja_long-term_2014,yang_using_2008}. 

The RF and CRNN classifiers are both promising candidates for deployment in a clinical system, achieving high accuracy scores and requiring low computational resources at test time in comparison to the k-NN and SVC algorithms. While the CRNN classifier achieved the best accuracy, RF required far less computational resources for training and scoring in comparison. The CRNN classifier is also attractive because by learning the sensor outputs directly, there is no need for hand coding, scaling, or selecting an optimal feature set, which simplifies and may improve generalizability of the ARC. 

As pointed out by \cite{ordonez_deep_2016}, the CRNN architecture applies intuitively to inertial sensor data. The input convolutional layers can be likened to a feature extractor, that responds maximally to specific temporal sequences within the timespan of the kernel and are learned directly from the dataset. The recurrent (LSTM) layers model the activation dynamics of the feature map. 

Sensor fusion performed by the Apple Watch operating system computes gravitational $\mathbf{g} = [g_x, g_y, g_z]$ and user $\mathbf{b} = [b_x,b_y,b_z]$ components of the total acceleration, which are also recorded by the PowerSense application. Although other authors \citep{gonzalez_features_2015, bulling_tutorial_2014} have utilized inertial sensor fusion for HAR, only raw sensor data is utilized in this study. Sensor fusion errors and response delays were apparent upon examining plots of the recorded data, and early experiments conducted incorporating body acceleration data did not improve classifier performance.

For cyclic activities, it has been demonstrated that window length selection represents a tradeoff between algorithm responsiveness and classifier performance \citep{banos_window_2014}, with increasing window length favoring higher classification accuracy. This is consistent with the results of the segmentation parameter experiments conducted in this study. A two second window length was found to be compatible with high classification accuracy and sufficiently responsive for the exercises and subjects in this study; where single exercise repetition time varied between one to four seconds. It is also notable that increasing segment overlap of the training data improved ARC performance monotonically, at the cost of increased computational resources during training. High training segmentation overlap also degrades k-NN computational and storage efficiency at test and deployment, but is likely to be a worthwhile strategy for boosting performance of the other classifiers in a clinical system. 

Better classification accuracy on unseen subjects may ultimately be achieved in future with training datasets incorporating more subjects. Another potential approach would be to utilize a 9-axis inertial sensor to break the inter-class symmetries about the gravitational axis. The 9-axis sensor magnetometer would provide a second absolute orientation reference perpendicular to the gravitational axis measured by the accelerometer, to achieve better class separation. Indeed,  \cite{shen_i_2016} demonstrated that a wrist-worn 9-axis sensor can be used to measure shoulder pose during active motion to within 10\degree\ accuracy. Disadvantages of using a 9-axis sensor, however, include the current limited selection of commercially availability smartwatches incorporating them, and the potential for signal disturbances from adjacent magnetic or electrically conductive objects.

The chief limitations of this study are the limited number of subjects (20) in the data set, and that no subjects had symptomatic shoulder disorders. Also our subjects were significantly younger than typical rotator cuff tear populations \citep{tashjian_epidemiology_2012}. How well the proposed ARC performance would generalize to a clinical population is uncertain. Furthermore, as the data was acquired during a single supervised session for each subject, it is uncertain how well algorithm performance would generalize to subsequent unsupervised sessions where performance of the exercise may change due to absence of supervision, or perhaps gradual improvement associated with shoulder recovery. 

\section{Conclusions}
This proof-of concept study demonstrates the feasibility of applying machine learning to wrist worn inertial sensor data for shoulder physiotherapy exercise recognition. This is an important step toward objective measurement of adherence to at-home shoulder physiotherapy exercise protocols. Future work will focus on translation of this technology to the clinical setting and evaluating exercise classification in shoulder disorder populations.

\section{Acknowledgments}
No potential conflict of interest or external funding source was reported by the authors.

\bibliographystyle{tfcse}
\bibliography{ShoulderSmartwatch}

\newpage

\section{Appendix}

\begin{figure}[h]
  \begin{center}
    \includegraphics[scale=0.38]{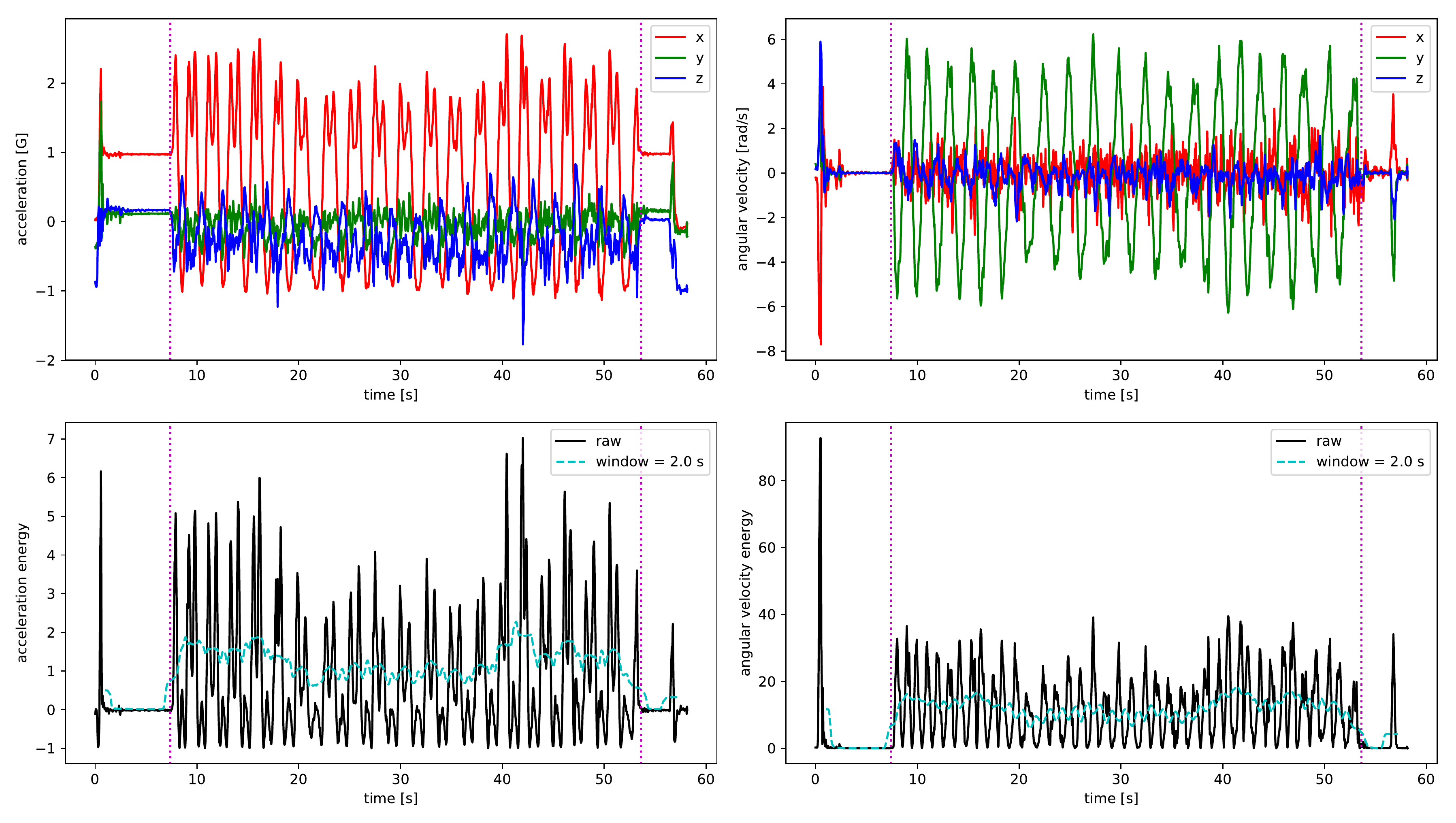}
  \end{center}
  \caption{Forward elevation raw data sample with ground truth annotation between vertical bars using 33\% threshold for 2s moving average acceleration energy filter.}
  \label{f:seg}
\end{figure}

\begin{figure}[]
\centering
\subfloat[Random forest]{%
\resizebox*{4.5cm}{!}{\includegraphics{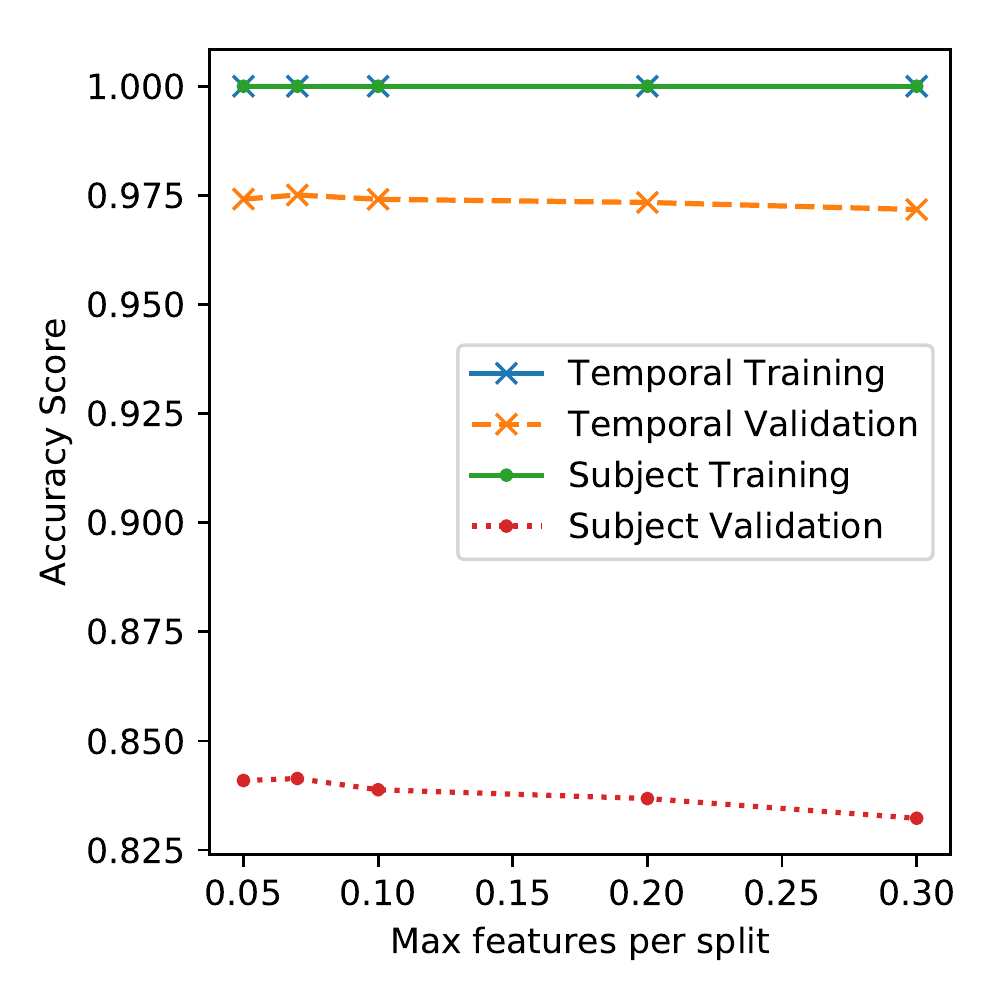}}}
\subfloat[k-NN]{%
\resizebox*{4.5cm}{!}{\includegraphics{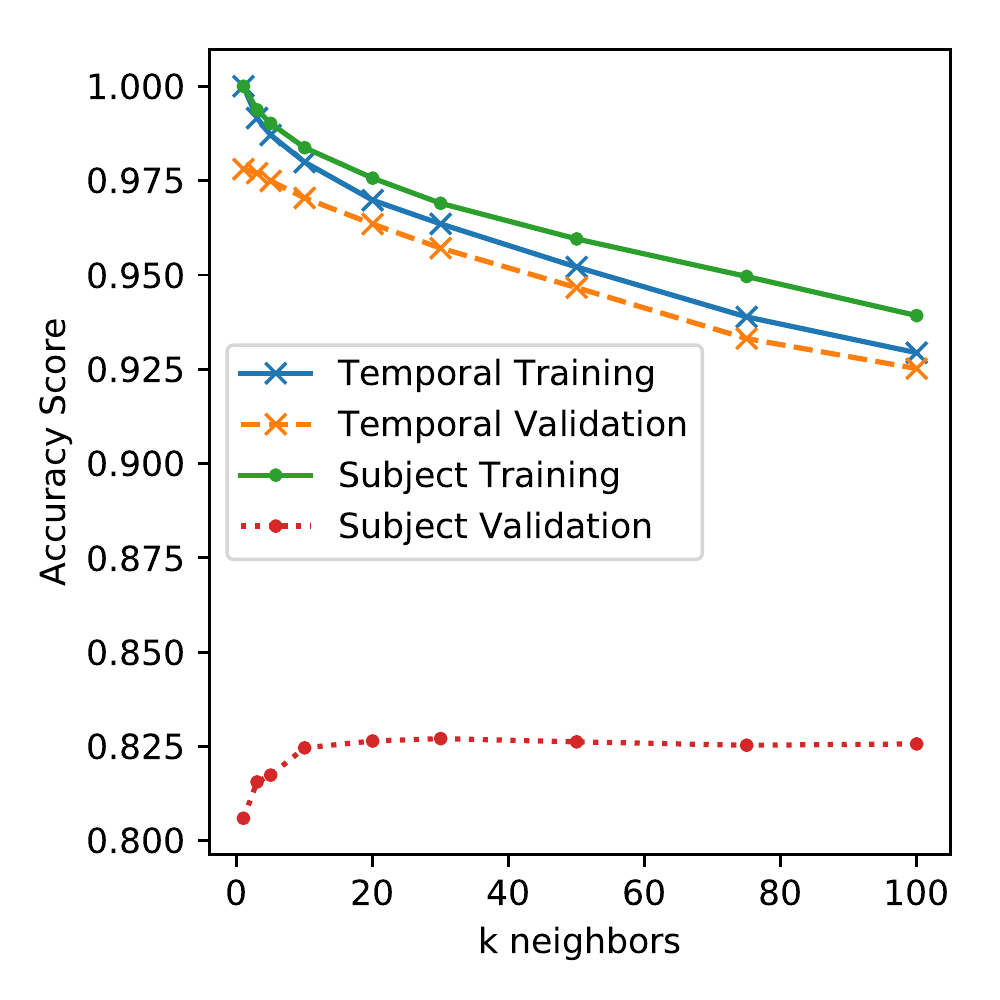}}}
\subfloat[SVC]{%
\resizebox*{4.5cm}{!}{\includegraphics{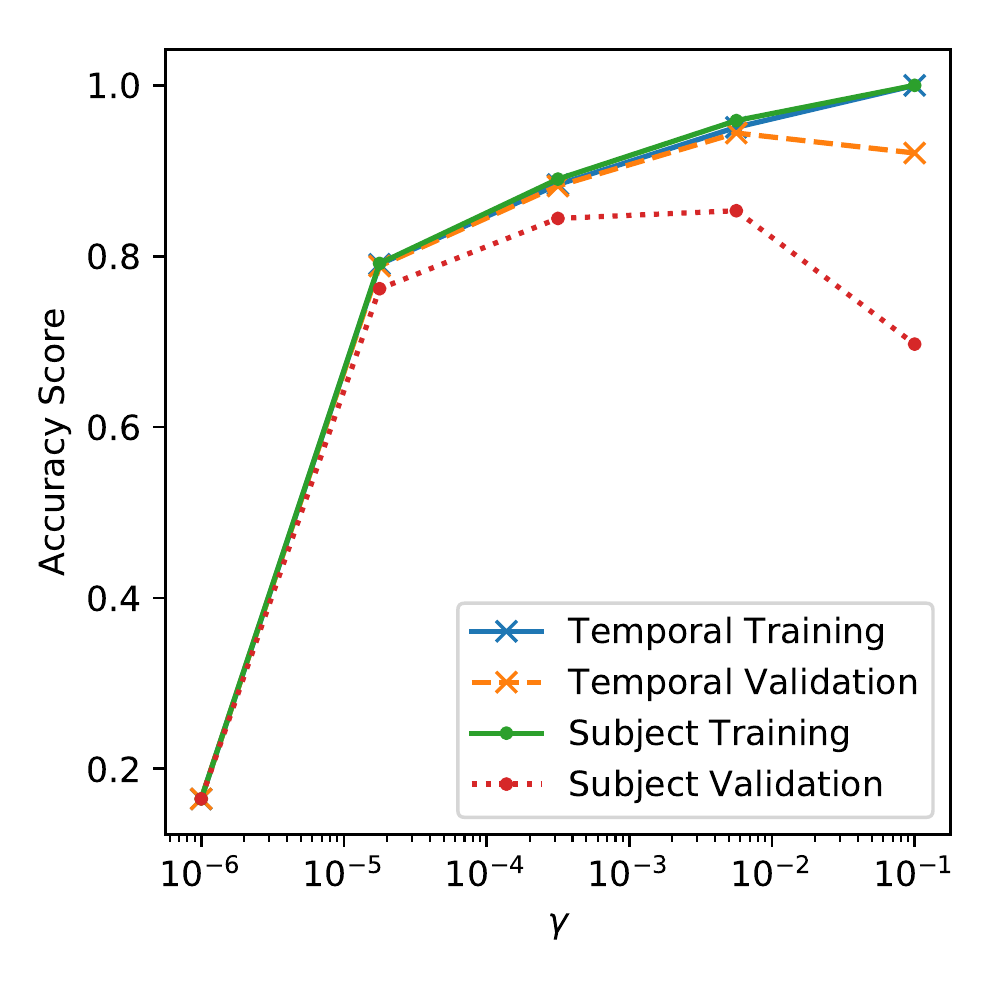}}}
\caption{Hyperparameter optimization of \textbf{(a)} Random forest maximum features per split, \textbf{(b)} k-NN k neighbors, \textbf{(c)} SVC kernel coefficient $\gamma$.} \label{f:hopt}
\end{figure}

\end{document}